\documentclass[a4paper,11pt]{article}
\usepackage[dvips]{graphicx}
\usepackage{amssymb}
\usepackage{amsmath}

\usepackage{cite}

\input arrow

\begin{document}

\title{Low Dimensional Supersymmetries in SUSY Chern-Simons Systems and Geometrical Implications}
\author{V. K. Oikonomou\thanks{
voiko@physics.auth.gr}\\
Max Planck Institute for Mathematics in the Sciences\\
Inselstrasse 22, 04103 Leipzig, Germany} \maketitle

\begin{abstract}
We study in detail the underlying graded geometric structure of abelian
$N=2$ supersymmetric Chern-Simons theory in $(2+1)$-dimensions.
This structure is an attribute of the hidden unbroken one
dimensional $N=2$ supersymmetries that the system also possesses.
We establish the result that the geometric structures
corresponding to the bosonic and to the fermionic sectors are
equivalent fibre bundles over the $(2+1)$-dimensional manifold. Moreover, we find a geometrical answer to the question
why some and not all of the fermionic sections are related to a $N=2$ supersymmetric algebra.
Our findings are useful for the
quantum theory of Chern-Simons vortices.
\end{abstract}

\section*{Introduction}

Chern-Simons terms \cite{lee1,lee1a,lee1b} are important terms that can be consistently be added to $(2+1)$-dimensional
Lagrangians of gauge field theories \cite{dunnebook,lee}. These
terms provide particularly interesting properties to the these
theories, altering the long distance behavior and modifying the
solitonic solutions \cite{lee1,lee1a,lee1b}.

\noindent The supersymmetric extensions of $(2+1)$-dimensional
gauged Chern-Simons models and particularly the ones with global
$N=2$ spacetime supersymmetry, have the interesting attribute which is the fact that
the zero modes of fermions and bosons are directly related to the
zero modes of the bosonic field fluctuations, with the latter
describing massless bosonic modes around the vortices (see
\cite{lee}, and references therein).

\noindent A complete quantum theory of supersymmetric Chern-Simons
vortices is still lacking, hence providing information related to
such theories may shed light to various aspects of this quantum
theory. Particularly, finding underlying hidden symmetries or
inherent patterns that can be classified according to an already
known symmetry, may reduce the complexity of a generally difficult
problem.

\noindent In this context, in a recent paper \cite{workinprogress} we presented an appealing property
of the zero modes of Chern-Simons models with global $N=2$ spacetime
supersymmetry. Specifically, we established fact that the
fermionic and bosonic zero modes of $N=2$ Abelian Chern-Simons gauge models in
$(2+1)$-dimensions are related to two $N=2$, $d=1$ supersymmetric
quantum algebras
\cite{witten,susyqm,susyqm1,susyqm2,susyqm3,susyqm4,susyqm5,susyqmarxiv,susyqmarxiv1,susyqmarxiv2,susyqmarxiv3,susyqmarxiv4,thaller,pluskai,plu1,plu2,plu3,plu4}.
We focused in the limiting case $\kappa =0$, with
$\kappa$ the coupling of the Chern-Simons term. The latter theory is
nothing else but a supersymmetric extension of the Landau-Ginzburg
model. As we demonstrated in \cite{workinprogress}, the two $N=2$, $d=1$, supersymmetries combine to form an $N=4$ extended supersymmetry, with non zero central charge \cite{ivanovbook,ivanov1,ivanov2,ivanov3,ivanov4,ivanov5,ivanov6,ivanov7,ivanov8,ivanov9,ivanov10,ivanov11}. We argued that this result is a direct consequence of the $N=2$ global spacetime supersymmetry, possessed by the initial system. To this end, in this paper we shall investigate the geometric implications that each $N=2$, $d=1$, supersymmetry generates and in addition we shall see that there is a geometric reason behind the fact that the one dimensional supersymmetries combine to form  $N=4$ extended supersymmetry.  

\noindent In particular, as we shall evince, the supersymmetric quantum mechanics (SUSY QM hereafter) algebras
provide a grading on the Hilbert space of solutions. In turn this
grading provides the system with an interesting underlying
geometric structure. As we shall see, the grading of the unbroken
$N=2$ SUSY QM algebra makes the $(2+1)$-dimensional space a graded
manifold, denoted $(X,\mathcal{A})$, with body $X$ and structure
sheaf $\mathcal{A}$. This graded manifold can be used to construct
composite fibre bundles over the space $X$. Regarding the fermionic
sector, the supersymmetric structure is only for some sections of
the spin bundle over $X$. With the help of the graded manifold we
can explain how this result occurs, since the covariant
differential of the $U(1)$-twisted spin bundle over $X$ is
reducible to a covariant differential of a composite fibre bundle
over $X$, a bundle that contains the graded bundle in its
structure. This reducibility is actually done in terms of the
corresponding connections of the composite fibre bundles. In the
bosonic case, the covariant differential of the corresponding
composite fibre bundle is projectable to the one of the total
$U(1)$-twisted fibre bundle, the sections of which are the bosonic
fluctuations. The fact that the fermionic and bosonic zero modes
of $N=2$ Abelian gauge models in $(2+1)$-dimensions are directly
related, has an immediate impact on the underlying geometric
structure. Particularly we shall establish the result that the
composite fibre bundles for fermions and bosons are equivalent,
which means that there is some isomorphism between these. This in
turn can be used to establish the fact that the bundles belong to
the same K-theoretic equivalence class. This K-theoretic equivalence is actually the reason that the two $N=2$, $d=1$ supersymmetries form an $N=4$, $d=1$ supersymmetry.

\noindent This paper is organized as follows. In section 1 we review the general theoretical framework of abelian $N=2$
supersymmetric Chern-Simons theory in $(2+1)$-dimensions and
briefly present the result of \cite{workinprogress},  that there is an unbroken $N=2$ SUSY QM
algebra underlying both the fermionic and bosonic sector of the
theory. In section 2 we describe in detail the additional
geometric structure over the $(2+1)$-dimensional space $X$, which
is implied by the unbroken $N=2$ SUSY QM algebra and discuss the
features of the fermionic and bosonic geometric structures over
$X$. We also discuss why these structures are equivalent, a
feature that must be a result of the dimensionality of the space
$X$ and also due to the global $N=2$ supersymmetry.

\section{Supersymmetric Abelian Chern-Simons Theory}

In order to make the article self-contained, we briefly review the abelian $N=2$ supersymmetric Chern-Simons theory in $(2+1)$-dimensions
and present the basic result of \cite{workinprogress}. As we already noted, one feature of the Abelian Chern-Simons supersymmetric model under study is that, within the $N=2$ supersymmetry framework, the
fermionic zero modes are directly related to the bosonic zero
modes. This fact has particularly interesting consequences, in
reference to the underlying one dimensional supersymmetries and
also for the geometrical structures, as we shall demonstrate in a later section. Let us
briefly present the models that we will use in the following.
Adopting the notation of references \cite{lee,workinprogress}, the Lagrangian for
the $N=2$ spacetime supersymmetric model is,
\begin{align}\label{n2lag}
&\mathcal{L}=-\frac{1}{4}F_{\mu \nu}^{\mu \nu}+\frac{1}{4}\kappa
\epsilon^{\mu \nu \lambda}F_{\mu \nu}A_{\lambda}-\lvert
D_{\mu}\phi \lvert^{2}-\frac{1}{2}(\partial_{\mu}N)^2
\\ \notag & -\frac{1}{2}(e\lvert\phi\lvert^2+\kappa N-eu^2)^2-e^2N^2\lvert\phi\lvert^2
\\ \notag & i\bar{\psi}\gamma^{\mu}D_{\mu}\psi+i\bar{\chi}\gamma^{\mu}\partial_{\mu}\chi+\kappa\bar{\chi}\chi
+i\sqrt{2}e(\bar{\psi}\chi\phi-\bar{\chi}\psi\phi^*)+e
N\bar{\psi}\psi
\end{align}
where $D_{\mu}=\partial_{\mu}-ieA_{\mu}$ and the gamma matrices
satisfy the relation
$\gamma^{\mu}\gamma^{\nu}=-\eta^{\mu\nu}-i\epsilon^{\mu \nu
\lambda}\gamma_{\lambda}$. The fields $\psi ,\chi$ are  two
component spinors, with $\psi$ being a Weyl charged fermionic
field and $\chi$ a neutral complex Weyl two component spinor.

\subsection{Zero Modes of the Fermionic Sector}

Since the zero modes of the fermionic sector of the $N=2$ Lagrangian in
the background of self-dual vortices are very important in what
follows, so let us see how we can find these. The fermionic
equations of motion corresponding to Lagrangian (\ref{n2lag}), are
given by the following equations:
\begin{align}\label{eqnmotion}
& \gamma^i D_i \psi+ie(\gamma^0 A^0-N)\psi-\sqrt{2}e\phi \chi=0
\\ \notag & \gamma^i\partial_i \chi -i\kappa \chi+\sqrt{2}e\phi^*\psi=0
\end{align}
By making the following conventions:
\begin{equation}\label{conventions}
\gamma^0=\sigma_3,{\,}{\,}\gamma^1=i\sigma_2,{\,}{\,}\gamma^2=i\sigma_1,{\,}{\,}{\,}
\psi=\left (\begin{array}{c}
        \psi_{\uparrow}          \\
    \psi_{\downarrow} \\
\end{array}\right ),{\,}{\,}{\,}\chi=\left (\begin{array}{c}
        \chi_{\uparrow}          \\
    \chi_{\downarrow} \\
\end{array}\right )
\end{equation}
and moreover assuming positive values of the flux, we obtain the
equations of motion for the fermions:
\begin{align}\label{fermionsequationsofmotions}
&(D_1+iD_2)\psi_{\downarrow}-\sqrt{2}e\phi \chi_{\uparrow}=0
\\ \notag & (\partial_1-i\partial_2)\chi_{\uparrow }+i\kappa \chi_{\downarrow }-\sqrt{2}\phi^*\psi_{\downarrow }=0
\\ \notag & (D_1-iD_2)\psi_{\uparrow }+2i e A^0\psi_{\downarrow }+\sqrt{2}e\phi \chi_{\downarrow }=0
\\ \notag & (\partial_1+i\partial_2)\chi_{\downarrow }-i\kappa \chi_{\uparrow }+\sqrt{2}e\phi^*\psi_{\uparrow }=0
\end{align}
We shall explore the theory for the limiting case $\kappa =0$ of
the Chern-Simons parameter $\kappa$. Note that, when the Chern
Simons coupling $\kappa$ becomes zero, the $N=2$ Lagrangian reduce
to the $N=2$ Abelian Higgs model.

\subsection{Fluctuating Bosonic Zero Modes of the $N=2$ Supersymmetric System}

In addition to the fermionic zero modes, the zero modes of the
bosonic fields fluctuations are very necessary to reveal the
underlying geometric and symmetric structure of the system. Let us
give a brief account on how to find these. In order to do this, we
focus our interest on the bosonic part of the Lagrangian
(\ref{n2lag}). The theory has two ground states, namely the
non-symmetric one, with $\lvert \phi \lvert =v$, $N=0$ and a
symmetric one with $\phi =0,{\,}{\,}N=\frac{ev^2}{\kappa}$.
Solutions of the topological soliton type exist in non-symmetric
phase which has the following asymptotic behavior \cite{lee}:
\begin{equation}\label{boun1}
\lim_{r\rightarrow \infty} N(r)\rightarrow
0,{\,}{\,}{\,}\lim_{r\rightarrow \infty}\lvert \phi (r)\lvert
\rightarrow v
\end{equation}
and additionally a quantized flux $\Phi=\pm \frac{2\pi n}{e}$. In
the symmetric Higgs phase, $\phi =0$, $N=ev^2/\kappa$,
non-topological solutions exist with the following asymptotic
behavior:
\begin{equation}\label{boun2}
\lim_{r\rightarrow \infty}N(r)\rightarrow
\frac{ev^2}{\kappa}+\frac{\mathrm{const.}}{r^{2a}},{\,}{\,}{\,}\lim_{r\rightarrow
\infty}\lvert \phi (r)\lvert \rightarrow
\frac{\mathrm{const.}}{r^a}
\end{equation}
Furthermore, all static solutions must satisfy the Gauss law:
\begin{equation}\label{gausslaw}
\partial_{i}F^{i0}+\kappa F_{12}-ie(\phi^*D^0\phi-D^0\phi^*\phi)=0
\end{equation}
Integrating over the whole space we obtain a static configuration
of magnetic flux $\Phi=\int \mathrm{d}^2xF_{12}$ which has a total
electric charge, $Q=-\frac{\kappa \Phi}{e}$. The energy of the
configuration is bounded from below by the relation $E\geq
ev^2\lvert \Phi\lvert $, and is saturated if the configurations
satisfy the self-duality equations:
\begin{align}\label{selfdualeqns}
&(D_1\pm iD_2)\phi=0
\\ \notag & F_{12}\pm(e\lvert \phi\lvert^2 +\kappa N-ev^2)=0
\\ \notag & A^0\mp N=0
\\ \notag & \partial_{i}F^{i0}+\kappa F_{12}-ie(\phi^*D^0\phi-D^0\phi^*\phi)=0
\end{align}

\noindent The equations of the zero modes fluctuations are
obtained by varying the self-duality equations (\ref{selfdualeqns}) around the static
classical vortex configuration, and can be cast in the following
form:
\begin{align}\label{selfdualfluctuat}
&(D_1+iD_2)\delta \phi-ie\phi (\delta A_1+i \delta A_2)=0
\\ \notag &\partial_1\delta A_2-\partial_2 \delta A_1+e(\phi^*\delta \phi +\phi\delta \phi^*)+k\delta A^0=0
\end{align}
We will set $\kappa =0$ in order to describe the Landau-Ginzburg
vortex situation, as we did in the fermion case. In that case, one
can consistently set $A^0=N=0$, just as we did in the fermionic
case.

\subsection{$N=2$ Supersymmetric Quantum Mechanics Algebra in the Fermionic Sector}

The fermionic equations of motion
(\ref{fermionsequationsofmotions}) for $\kappa =0$ can be cast as,
\begin{align}\label{fer1}
&(D_1+iD_2)\psi_{\downarrow}-\sqrt{2}e\phi\chi_{\uparrow}=0
\\ \notag & (\partial_1-i\partial_2)\chi_{\uparrow}-\sqrt{2}e\phi^*\psi_{\downarrow}=0
\\ \notag & (D_1-iD_2)\psi_{\uparrow}+\sqrt{2}e\phi\chi_{\downarrow}=0
\\ \notag & (\partial_1+i\partial_2)\chi_{\downarrow}+\sqrt{2}e\phi^*\psi_{\uparrow}=0
\end{align}
The last two equations of relation (\ref{fer1}) have no solutions
describing localized fermions, but only have some trivial
solutions \cite{lee}. However, the first two equations of relation
(\ref{fer1}), have $2n$ normalized solutions, with $n$ the vorticity number
\cite{lee}. Thereby, we can form the operator $\mathcal{D}_{LG}$,
corresponding to the first two equations of (\ref{fer1}),
\begin{equation}\label{susyqmrn567m}
\mathcal{D}_{LG}=\left(%
\begin{array}{cc}
 D_1+iD_2 & -\sqrt{2}e\phi
 \\ -\sqrt{2}\phi^* & \partial_1-i\partial_2\\
\end{array}%
\right)
\end{equation}
acting on the vector:
\begin{equation}\label{ait34e1}
|\Psi_{LG}\rangle =\left(%
\begin{array}{c}
  \psi_{\downarrow} \\
  \chi_{\uparrow} \\
\end{array}%
\right).
\end{equation}
Consequently, the first two equations of (\ref{fer1}) can be cast
as:
\begin{equation}\label{transf}
\mathcal{D}_{LG}|\Psi_{LG}\rangle=0
\end{equation}
The solutions of the above equation are the zero modes of the
operator $\mathcal{D}_{LG}$. Recalling that the first two
equations of (\ref{fer1}) have $2n$ normalized solutions, we can
easily state that:
\begin{equation}\label{dimeker}
\mathrm{dim}{\,}\mathrm{ker}\mathcal{D}_{LG}=2n
\end{equation}
Furthermore, the adjoint of the operator $\mathcal{D}_{LG}$,
namely $\mathcal{D}_{LG}^{\dag}$, is equal to:
\begin{equation}\label{eqndag}
\mathcal{D}_{LG}^{\dag}=\left(%
\begin{array}{cc}
 D_1-iD_2 & \sqrt{2}e\phi
 \\ \sqrt{2}\phi^* & \partial_1+i\partial_2\\
\end{array}%
\right)
\end{equation}
and acts on the vector:
\begin{equation}\label{ait3hgjhgj4e1}
|\Psi_{LG}'\rangle =\left(%
\begin{array}{c}
  \psi_{\uparrow} \\
  \chi_{\downarrow} \\
\end{array}%
\right).
\end{equation}
The zero modes of the adjoint operator $\mathcal{D}_{LG}^{\dag}$,
correspond to the solutions of the last two equations of
(\ref{fer1}), with the obvious replacement $e\rightarrow-e$.
Obviously, since the last pair of equations of relation
(\ref{fer1}) have no normalized solutions, the corresponding
kernel of the adjoint operator is null, that is:
\begin{equation}\label{dimeke1r11}
\mathrm{dim}{\,}\mathrm{ker}\mathcal{D}_{LG}^{\dag}=0
\end{equation}
The normalization condition for the solutions of (\ref{fer1}) is
crucial for our analysis, since only for such solutions the
operator $\mathcal{D}_{LG}$ is Fredholm, a result that can be
verified by (\ref{dimeker}) and (\ref{dimeke1r11}). The fermionic
system in the self-dual vortices background, possesses an unbroken
$N=2$, $d=1$ supersymmetry. Indeed, we can form the supercharges
and the quantum Hamiltonian of this $N=2$, $d=1$ SUSY algebra in
terms of the operator $\mathcal{D}_{LG}$, and these are equal to:
\begin{equation}\label{s7}
\mathcal{Q}_{LG}=\bigg{(}\begin{array}{ccc}
  0 & \mathcal{D}_{LG} \\
  0 & 0  \\
\end{array}\bigg{)},{\,}{\,}{\,}\mathcal{Q}^{\dag}_{LG}=\bigg{(}\begin{array}{ccc}
  0 & 0 \\
  \mathcal{D}_{LG}^{\dag} & 0  \\
\end{array}\bigg{)},{\,}{\,}{\,}\mathcal{H}_{LG}=\bigg{(}\begin{array}{ccc}
 \mathcal{D}_{LG}\mathcal{D}_{LG}^{\dag} & 0 \\
  0 & \mathcal{D}_{LG}^{\dag}\mathcal{D}_{LG}  \\
\end{array}\bigg{)}
\end{equation}
These three elements of the algebra, satisfy the $d=1$ SUSY QM
algebra:
\begin{equation}\label{relationsforsusy}
\{\mathcal{Q}_{LG},\mathcal{Q}^{\dag}_{LG}\}=\mathcal{H}_{LG}{\,}{\,},\mathcal{Q}_{LG}^2=0,{\,}{\,}{\mathcal{Q}_{LG}^{\dag}}^2=0
\end{equation}
The Hilbert space of the supersymmetric quantum mechanical system,
$\mathcal{H}_{LG}$ is an $Z_2$ graded vector space, with the
grading provided by the operator $\mathcal{W}$, an involution
operator known as the Witten parity. This operator commutes with
the total Hamiltonian and anti-commutes with the supercharges,
\begin{equation}\label{s45}
[\mathcal{W},\mathcal{H}_{LG}]=0,{\,}{\,}\{\mathcal{W},\mathcal{Q}_{LG}\}=\{\mathcal{W},\mathcal{Q}_{LG}^{\dag}\}=0
\end{equation}
Moreover, the Witten parity $\mathcal{W}$, satisfies the following
identity,
\begin{equation}\label{s6}
\mathcal{W}^{2}=1
\end{equation}
As we already mentioned, the Witten parity $\mathcal{W}$, spans
the total Hilbert space into subspaces which are classified
according to their $Z_2$ parity. Hence the total Hilbert space of
the quantum system can be written as:
\begin{equation}\label{fgjhil}
\mathcal{H}=\mathcal{H}^+\oplus \mathcal{H}^-
\end{equation}
with the vectors that belong to the two subspaces
$\mathcal{H}^{\pm}$, classified according to their Witten parity,
to even and odd parity states, that is:
\begin{equation}\label{shoes}
\mathcal{H}^{\pm}=\mathcal{P}^{\pm}\mathcal{H}=\{|\psi\rangle :
\mathcal{W}|\psi\rangle=\pm |\psi\rangle \}
\end{equation}
Furthermore, the corresponding Hamiltonians of the $Z_2$ graded
spaces are:
\begin{equation}\label{h1}
{\mathcal{H}}_{+}=\mathcal{D}_{LG}{\,}\mathcal{D}_{LG}^{\dag},{\,}{\,}{\,}{\,}{\,}{\,}{\,}{\mathcal{H}}_{-}=\mathcal{D}_{LG}^{\dag}{\,}\mathcal{D}_{LG}
\end{equation}
The operator $\mathcal{W}$, in the case at hand, can be
represented in the following matrix form:
\begin{equation}\label{wittndrf}
\mathcal{W}=\bigg{(}\begin{array}{ccc}
  1 & 0 \\
  0 & -1  \\
\end{array}\bigg{)}
\end{equation}
The eigenstates of $\mathcal{P}^{\pm}$ which are denoted
$|\psi^{\pm}\rangle$, satisfy the following relation:
\begin{equation}\label{fd1}
P^{\pm}|\psi^{\pm}\rangle =\pm |\psi^{\pm}\rangle
\end{equation}
Hence, we call them positive and negative parity eigenstates, with
``parity'' referring to the $P^{\pm}$ operator, which is nothing
else but the Witten parity operator. Using the representation
(\ref{wittndrf}) for the Witten parity operator,
 the parity eigenstates can represented by the vectors,
\begin{equation}\label{phi5}
|\psi^{+}\rangle =\left(%
\begin{array}{c}
  |\phi^{+}\rangle \\
  0 \\
\end{array}%
\right),{\,}{\,}{\,}
|\psi^{-}\rangle =\left(%
\begin{array}{c}
  0 \\
  |\phi^{-}\rangle \\
\end{array}%
\right)
\end{equation}
with $|\phi^{\pm}\rangle$ $\in$ $\mathcal{H}^{\pm}$. Turning
back to the fermionic system at hand, it is easy to write the
fermionic states of the system in terms of the SUSY QM algebra. We
already wrote down the supercharges, namely relation (\ref{s7}),
and then it is easy to verify that:
\begin{equation}\label{fdgdfgh}
|\Psi_{LG}\rangle =|\phi^{-}\rangle=\left(%
\begin{array}{c}
  \psi_{\downarrow} \\
  \chi_{\uparrow} \\
\end{array}%
\right),{\,}{\,}{\,}|\Psi_{LG}'\rangle =|\phi^{+}\rangle=\left(%
\begin{array}{c}
  \psi_{\uparrow} \\
  \chi_{\downarrow} \\
\end{array}%
\right)
\end{equation}
Therefore, the corresponding even and odd parity SUSY QM states
are the following:
\begin{equation}\label{phi5}
|\psi^{+}\rangle =\left(%
\begin{array}{c}
  |\Psi_{LG}'\rangle \\
  0 \\
\end{array}%
\right),{\,}{\,}{\,}
|\psi^{-}\rangle =\left(%
\begin{array}{c}
  0 \\
  |\Psi_{LG}\rangle \\
\end{array}%
\right)
\end{equation}
on which, the Hamiltonian and the supercharges act. Supersymmetry
is unbroken if the Witten index is a non-zero integer. The Witten
index for Fredholm operators is equal to:
\begin{equation}\label{phil}
\Delta =n_{-}-n_{+}
\end{equation}
with $n_{\pm}$ the number of zero modes of ${\mathcal{H}}_{\pm}$
in the subspace $\mathcal{H}^{\pm}$, with the constraint that
these are finitely many.

\noindent When the Witten index is zero and also if
$n_{+}=n_{-}=0$, then supersymmetry is broken. However, if $n_{+}=
n_{-}\neq 0$ the system has still an unbroken supersymmetry.

\noindent The Witten index is connected to the Fredholm index of
the operator $\mathcal{D}_{LG}$, as follows:
\begin{align}\label{ker1}
&\Delta=\mathrm{dim}{\,}\mathrm{ker}
{\mathcal{H}}_{-}-\mathrm{dim}{\,}\mathrm{ker} {\mathcal{H}}_{+}=
\mathrm{dim}{\,}\mathrm{ker}\mathcal{D}_{LG}^{\dag}\mathcal{D}_{LG}-\mathrm{dim}{\,}\mathrm{ker}\mathcal{D}_{LG}\mathcal{D}_{LG}^{\dag}=
\\ \notag & \mathrm{ind} \mathcal{D}_{LG} = \mathrm{dim}{\,}\mathrm{ker}
\mathcal{D}_{LG}-\mathrm{dim}{\,}\mathrm{ker}
\mathcal{D}_{LG}^{\dag}
\end{align}
Using equations (\ref{dimeker}) and (\ref{dimeke1r11}), the Witten
index is equal to:
\begin{equation}\label{fnwitten}
\Delta =-2n
\end{equation}
Thereupon, the fermionic system in the self-dual Landau-Ginzburg
vortices background with $N=2$ spacetime supersymmetry, has an
$N=2$, $d=1$ unbroken supersymmetry. We could argue that this
result could stem from the fact that the initial system has an
unbroken $N=2$ spacetime supersymmetry, so the Hilbert space of
the zero modes states also has a remnant $N=2$, $d=1$
supersymmetric quantum algebra. However, this is not true since
global spacetime supersymmetry in $d>1$ dimensions and
supersymmetric quantum mechanics, that is $d=1$ supersymmetry, are
not the same. The SUSY QM supercharges do not generate
transformations between fermions and bosons and also these
supercharges classify the Hilbert space of quantum states
according to the group $Z_2$. So we can state that the
supersymmetric quantum mechanics algebra is not similar to a
global spacetime supersymmetry, but nevertheless the SUSY QM
algebra is a graded algebra.

\subsection{$N=2$ Supersymmetric Quantum Mechanics Algebra of the Bosonic Fluctuations}

We now turn our focus to the bosonic zero modes equations. For
$\kappa =0$ and $A^0=N=0$ these can be cast in the following form:
\begin{align}\label{selfdualfluctuat1fd}
&(D_1+iD_2)\delta \phi-ie\phi (\delta A_1+i \delta A_2)=0
\\ \notag &(\partial_1-\partial_2)(\delta A_1+ i\delta A_2)+2ie\phi^*\delta \phi =0
\end{align}
Interestingly enough, the above equations become identical to
equations (\ref{selfdualeqns}), if we substitute:
\begin{equation}\label{biossubs}
\psi_{\downarrow}=\delta
\phi,{\,}{\,}{\,}\chi_{\uparrow}=\frac{i}{\sqrt{2}}(\delta
A_1+i\delta A_2)
\end{equation}
As we shall see, this will play a crucial role when we  will
address the equivalence of the geometrical structures
corresponding to fermions and bosons. Due to this fact, it is easy
to conclude that the number of the zero modes corresponding to
equation (\ref{selfdualfluctuat1fd}), are equal to the total number
of zero modes corresponding to the first two equations of relation (\ref{fer1}), that
is $2n$. As in the fermionic case, a $N=2$ SUSY quantum mechanical
algebra underlies the bosonic system as well. The structure of the
algebra is the same as the fermionic one, with the difference that
the corresponding Hilbert space vectors are different. Indeed,
equations (\ref{selfdualfluctuat1fd}) can be written in the
following form:
\begin{equation}\label{transfrey}
\mathcal{D}_{LG}'|\Phi_{LG}\rangle=0
\end{equation}
where, the operator $\mathcal{D}_{LG}'$ is:
\begin{equation}\label{susyqmrrtyurn567m}
\mathcal{D}_{LG}'=\left(%
\begin{array}{cc}
 D_1+iD_2 & -\sqrt{2}e\phi
 \\ -\sqrt{2}\phi^* & \partial_1-i\partial_2\\
\end{array}%
\right)
\end{equation}
and acts on the vector:
\begin{equation}\label{ait3urtu4e1}
|\Phi_{LG}\rangle =\left(%
\begin{array}{c}
  \delta \phi \\
  \frac{i}{\sqrt{2}}(\delta A_1+i\delta A_2) \\
\end{array}%
\right).
\end{equation}
Hence we arrive to the conclusion that:
\begin{equation}\label{dimektriier}
\mathrm{dim}{\,}\mathrm{ker}\mathcal{D}_{LG}'=2n
\end{equation}
Furthermore, as in the fermionic case, the adjoint
${\mathcal{D}_{LG}'}^{\dag}$ has null ``ker'',
\begin{equation}\label{dimekegkjl1r11}
\mathrm{dim}{\,}\mathrm{ker}{\mathcal{D}_{LG}'}^{\dag}=0
\end{equation}
The operators ${\mathcal{D}_{LG}}^{\dag}$ and
${\mathcal{D}_{LG}'}^{\dag}$ are Fredholm as well, and thereby any
which operator constructed from these, is also Fredholm. The
supercharges and the Hamiltonian, that constitute the $N=2$, $d=1$
algebra in the bosonic case, are:
\begin{equation}\label{sgggggg7}
\mathcal{Q}_{LG}'=\bigg{(}\begin{array}{ccc}
  0 & \mathcal{D}_{LG}' \\
  0 & 0  \\
\end{array}\bigg{)},{\,}{\,}{\,}{{\mathcal{Q}'}^{\dag}}_{LG}=\bigg{(}\begin{array}{ccc}
  0 & 0 \\
  {\mathcal{D}'}_{LG}^{\dag} & 0  \\
\end{array}\bigg{)},{\,}{\,}{\,}\mathcal{H}_{LG}'=\bigg{(}\begin{array}{ccc}
 \mathcal{D}_{LG}\mathcal{D}_{LG}^{\dag} & 0 \\
  0 & \mathcal{D}_{LG}^{\dag}\mathcal{D}_{LG}  \\
\end{array}\bigg{)}
\end{equation}

\noindent These three elements of the algebra, also satisfy the
$d=1$ SUSY algebra:
\begin{equation}\label{relationsforsusy}
\{\mathcal{Q}_{LG}',{\mathcal{Q}'}^{\dag}_{LG}\}=\mathcal{H}_{LG}'{\,}{\,},\mathcal{Q}_{LG}^2=0,{\,}{\,}{\mathcal{Q}_{LG}^{\dag}}^2=0
\end{equation}
Supersymmetry is unbroken, since the corresponding Witten index
$\Delta '$ is a non-zero integer, in this case too. Indeed:
\begin{equation}\label{wittindexfgo}
\Delta '= -2n
\end{equation}
We found that the bosonic fluctuations in the self-dual
Landau-Ginzburg vortex background, are related to an $N=2$ SUSY
quantum mechanics algebra, which is identical to the fermionic
SUSY quantum mechanics algebra, that we came across earlier. We
now turn our focus to find what impact has this SUSY QM algebra on
the geometric structures that can be constructed over the
$(2+1)$-dimensional space.  

\section{Geometrical Implications of the $N=2$ SUSY QM Algebras}

The existence of underlying $N=2$, $d=1$ supersymmetric quantum algebras
in the fermionic and bosonic systems we presented in the previous
sections, has some geometrical implications on the geometric
spaces over $X$, to which spaces the fermions and bosons are sections of the corresponding fibre bundles.   

\noindent It is convenient to discuss in short what is our aim, why we are motivated to
search to find such extra geometric structures over $X$ and also what are our main results. The motivation
to search for extra underlying geometrical structure comes from the fact that
both in the fermionic and bosonic sector, there exists an underlying graded Hilbert vector space, given by vectors of the 
form (\ref{fdgdfgh}) and also (\ref{ait3urtu4e1}). Since the fermions and bosons are sections of some total fibre bundles, it is obvious
that some of these sections belong to another affine vector bundle that has some sort of an inherent graded structure. We shall construct such a structure in the following. Moreover,
in reference to fermions, not all sections of the total spin bundle, belong to this graded manifold. This can be verified by looking relation (\ref{fer1}). Only the first
two of these equations are associated to an unbroken $N=2$ SUSY QM algebra. Therefore, this suggests that the connection of the total spin manifold,
is reducible to some other connection which is related to the underlying graded manifold. Finally, the fact that the operators
(\ref{susyqmrn567m}) and (\ref{susyqmrrtyurn567m}) for fermions and bosons are equal, strongly suggests some equivalence relation between the corresponding 
composite graded fiber structures. In the following subsections we shall extensively address
these issues.

\subsection{Geometric Structures for the Fermionic Sector}

We denote $X$ the $(2+1)$-dimensional spacetime upon which the
model we described in this paper is built on. We first study the
fermionic sector. The fermions are sections of the $U(1)-$twisted
fibre bundle $P\times S\otimes U(1)$, where $S$ is the
representation of the Spin group $Spin(3)$, which in three
dimensions is irreducible, and $P$, the double cover of the
principal $SO(3)$ bundle on the tangent manifold $TX$. Some of
these spinors belong to the vector space of the supersymmetric
quantum algebra. The graded vector space
$\mathcal{H}=\mathcal{H}^+\oplus \mathcal{H}^-$, that some of the
sections of the fibre bundle $P\times S\otimes U(1)$ belong,
implies a new structure on the manifold $X$ and particularly $X$
is up-lifted to a graded manifold $(X,\mathcal{A})$ (for the
issues of connections on manifolds see for example \cite{graded1,Jost,Nakahara,eguchi}
while for connections on graded manifolds see \cite{graded1,graded}).
Indeed, the existence of the $N=2$ SUSY QM algebra,
$\mathcal{W},\mathcal{Q}_{LG},{Q}_{LG}^{\dag}$ and especially the
involution $W$, generates the $Z_2$ graded vector space
$\mathcal{H}=\mathcal{H}^+\oplus \mathcal{H}^-$. The subset
$\mathcal{H}^+$ contains $W$-even vectors and $W$-odd vectors.
This grading in turn is an additional algebraic structure on the
$(2+1)$-dimensional manifold $X$. As we already mentioned, $X$ is
up-lifted to a graded manifold $(X,\mathcal{A})$, with
$\mathcal{A}=\mathcal{A}^+\oplus \mathcal{A}^-$ an $Z_2$ graded
algebra. Particularly, $\mathcal{A}$ is a sheaf of
$Z_2$-graded commutative $R$-algebras of total rank $m$ ($m=2$ for
our case). Moreover, the sheaf $\mathcal{A}$ underlies the vector
space $\mathcal{H}$ and this sheaf makes the space $\mathcal{H}$
an $Z_2$-graded $\mathcal{A}$-module. Indeed, this can be verified
from the fact that:
\begin{equation}\label{amodule}
A_+\cdot M_+\subset M_+,{\,}{\,}A_+\cdot M_-\subset
M_-,{\,}{\,}A_-\cdot M_+\subset M_-,{\,}{\,}A_-\cdot M_-\subset
M_+
\end{equation}
The sheaf $\mathcal{A}$ contains the endomorphism $W$ (the
involution of the SUSY quantum algebra), $W:\mathcal{H}\rightarrow
\mathcal{H}$, with $W^2=I$, which provides the $Z_2$-grading on
$\mathcal{H}$, {\it i.e.}:
\begin{equation}\label{gred}
W\mathcal{H}^{\pm}=\pm 1
\end{equation}
Hence, $\mathrm{End}(\mathcal{H})\subseteq \mathcal{A}$. The sheaf
$\mathcal{A}$ is called a structure sheaf of the graded manifold
$(X,\mathcal{A})$, while $X$ is called the body of
$(X,\mathcal{A})$. In the following we shall mainly be interested
on the connections of this graded manifold, hence it worths noting
that given an open neighborhood $U$ of $x$ $\in $ $X$, we have
that locally (recall that $m=2$):
\begin{equation}\label{sheafloc}
\mathcal{A}(U)=C^{\infty}(U)\otimes\wedge R^m
\end{equation}
Hence the structure sheaf $\mathcal{A}$ is isomorphic to the sheaf
$C^{\infty}(U)\otimes\wedge R^m$ of some exterior vector bundle
$\wedge \mathcal{H_E}^*=U\times \wedge R^m$, with $\mathcal{H_E}$
an affine vector bundle with fiber the vector space $\mathcal{H}$. Then
actually, the structure sheaf
$\mathcal{A}=C^{\infty}(U)\otimes\wedge \mathcal{H}$, is
isomorphic to the sheaf of sections of the exterior vector bundle
$\wedge \mathcal{H_E}^*=R\oplus
(\oplus^{m}_{k=1}\wedge^k)\mathcal{H_E}^*$. The fibre bundle
$\mathcal{H_E}$ is very important for the definition of a graded
connection.

\noindent The connections of the graded manifold $(X,\mathcal{A})$
constitute an affine space modeled on the linear space of
sections of the vector bundle $TX^*\otimes \wedge
\mathcal{H_E}^*\otimes \mathcal{H_E}$. Note that, a connection
preserves the grading of the vector space
$\mathcal{H}=\mathcal{H}^+\oplus \mathcal{H}^-$. The sections of
the bundle $TX^*\otimes \wedge \mathcal{H_E}^*\otimes
\mathcal{H_E}$ are operators that belong to the sheaf
$\mathcal{A}$ with $\mathcal{H}$-valued forms as elements in some
specific representation, the dimension of which depends on the
rank of the sheaf. In the case at hand, these are $2\times 2$
matrices with $\mathcal{H}$-valued forms as matrix
elements. The graded connection we just described will be used in
the following as an auxiliary object, since we are not interested
in this connection, but in the composition of this with
another connection we shall describe.

\noindent Recall that the Dirac covariant differential corresponding to
the connection of the total bundle, $P\times S\otimes U(1)$, which
we denote $\gamma_s$, results to the equations (\ref{fer1}).
The fact that some of these equations are associated to 
an unbroken $N=2$ SUSY QM algebra, implies that the connection $\gamma_s$ is reducible to
another connection $\gamma_C$, related in some way to the
graded manifold $(X,\mathcal{A})$. Actually this reducibility
implies that the total covariant differential of the fibre bundle
$P\times S\otimes U(1)$, reduces to a simpler one, which when
integral sections of the fibre bundle $P\times S\otimes U(1)$ are
taken into account, results to the first two equations of relation
(\ref{fer1}). In this article, we shall
predominantly be interested for integral sections of the fibre
bundles that we will use. The non-zero modes are also interesting
since these provide information for the quantum theory of supersymmetric
Chern-Simons vortices, as we saw in \cite{workinprogress}. The $N=2$ SUSY QM algebra implies an
underlying geometric structure that is described by the following
diagram:
$$
\harrowlength=130pt \varrowlength=60pt \sarrowlength=120pt
\commdiag{X&\mapright^{\gamma_s}&\mathcal{S}\times\mathcal{P}\otimes
U(1){\,}{\,}{\,}{\,}{\,}\cr
&\arrow(3,-2)\lft{\gamma_E}&\mapup\rt{\gamma_{SE}}\cr &&
{\,}{\,}{\,}TX^*\otimes \wedge \mathcal{H_E}^*\otimes
\mathcal{H_E}\cr}$$ The arrows do not show the direction of the
projective maps of the corresponding fibre bundles, but the
directions of the connections of the corresponding fibre bundles.
The connections appearing on the arrows are defined to be
morphisms of the following fibre maps:
\begin{align}\label{mapconn}
& \gamma_s: P\times S\otimes U(1)\rightarrow J^1(P\times S\otimes
U(1)),
\\ \notag&(\mathrm{Bundle}{\,}\mathrm{map},{\,}{\,}\pi_s:P\times S\otimes U(1)\rightarrow X)
\\ \notag
& \gamma_E: TX^*\otimes \wedge \mathcal{H_E}^*\otimes
\mathcal{H_E} \rightarrow J^1(TX^*\otimes \wedge
\mathcal{H_E}^*\otimes \mathcal{H_E})
\\ \notag & (\mathrm{Bundle}{\,}\mathrm{map},{\,}{\,}\pi_E:TX^*\otimes \wedge \mathcal{H_E}^*\otimes \mathcal{H_E}\rightarrow X) \\ \notag
& \gamma_{SE}: (P\times S\otimes U(1))_G\rightarrow J^1((P\times
S\otimes U(1))_G)
\\ \notag &(\mathrm{Bundle}{\,}\mathrm{map},{\,}{\,}\pi_{SE}: P\times S\otimes U(1)\rightarrow TX^*\otimes \wedge \mathcal{H_E}^*\otimes \mathcal{H_E})
\end{align}
with $J^1Y_i$ the jet bundle of the corresponding bundle $Y_i$.
The underlying geometrical structure is a composite fibre bundle
over the manifold $X$, which is:
\begin{equation}\label{compbund}
P\times S\otimes U(1)\xrightarrow{\pi_{SE}} TX^*\otimes \wedge
\mathcal{H_E}^*\otimes \mathcal{H_E} \xrightarrow{\pi_{E}} X
\end{equation}
We can define the composite connection corresponding to the
composite fibre bundle (\ref{compbund}) as follows:
\begin{equation}\label{compconnection}
\gamma_C=\gamma_{SE}\circ \gamma_E
\end{equation}
which is nothing else than the composition of the connections
$\gamma_{SE}$ and $\gamma_E$. In order such a connection to exist,
there must be a canonical map between the jet bundles of the fibre
bundles that constitute the composite fibre bundle. The composite
connections have a direct impact on the first order differential
operators and the corresponding covariant differentials. It is
convenient here to remember the definition of the first order
differential and of the covariant differential corresponding to
some connection. For a general fibre bundle $Y\rightarrow X$, a
section $s_Y: X\rightarrow Y$, and a connection $\gamma_Y:
Y\rightarrow J^1 Y$, the first order differential is:
\begin{equation}\label{fod}
\mathcal{D}_{\gamma_Y}:J^1 Y\rightarrow TX^*\otimes VY
\end{equation}
with $VY$ the vertical subbundle of $Y$, which in the case of $Y$
is a vector bundle $VY=Y\times Y$. The covariant differential
corresponding to $\gamma_Y$, denoted $\nabla_{\gamma_Y}$, is:
\begin{equation}\label{covdiff}
\nabla^{\gamma_Y}=\mathcal{D}_{\gamma_Y}\circ J^1 s_Y:
X\rightarrow TX^*\otimes VY
\end{equation}
Hence, the total covariant differential of the bundle $P\times
S\otimes U(1)\rightarrow X$, is equal to:
\begin{equation}\label{totcovdif}
\nabla^{\gamma_{s}}=\mathcal{D}_{\gamma_{s}}\circ J^1 s:
X\rightarrow TX^*\otimes V(P\times S\otimes U(1))\equiv
TX^*\otimes P\times S\otimes U(1)
\end{equation}
Note that this covariant differential, when it acts on integral sections, 
results to the set of equations
(\ref{fer1}). Before we proceed, let us discuss something
very crucial for the proceeding analysis. Let $s_E$ and $s_{SE}$
be the sections of the following bundles:
\begin{align}\label{mapconn}
& s_E:X\rightarrow TX^*\otimes \wedge \mathcal{H_E}^*\otimes
\mathcal{H_E}
\\ \notag
& s_{SE}: TX^*\otimes \wedge \mathcal{H_E}^*\otimes
\mathcal{H_E}\rightarrow P\times S\otimes U(1)
\end{align}
We denote $Y_h$, the restriction $Y_h=s_E^*(P\times S\otimes U(1))$ of the fibre
bundle $\pi_{SE}: P\times S\otimes U(1)\rightarrow TX^*\otimes
\wedge \mathcal{H_E}^*\otimes \mathcal{H_E}$, to the submanifold
$s_E(X)\subset P\times S\otimes U(1)$, through the inclusion
\begin{equation}\label{8inc}
i_h:Y_h \hookrightarrow P\times S\otimes U(1)
\end{equation}
Given the sections $s_E$ and $s_{SE}$, their composition is $s_C=s_{SE}\circ s_E$, which is
a section of the fibre bundle $\pi_s:P\times S\otimes
U(1)\rightarrow X$, with $s_C(X)\subset P\times S\otimes U(1)$.
The definition of the covariant differential for the composite
bundle, denoted as $\nabla^{\gamma_{C}}$, corresponding to the connection
(\ref{compconnection}) follows easily:
\begin{equation}\label{totcovdifcomp}
\nabla^{\gamma_{C}}=\mathcal{D}_{\gamma_{C}}\circ J^1 s_C:
X\rightarrow TX^*\otimes VY_h\equiv TX^*\otimes Y_h
\end{equation}
The covariant differential when applied to integral sections of
$Y_h$ (which are a subset of the integral sections of $P\times
S\otimes U(1)$), gives rise to the first two equations of relation
(\ref{fer1}). Therefore, we observe that the
covariant differential, $\nabla^{\gamma_{s}}$ of the total bundle
$P\times S\otimes U(1)$ is reducible to the covariant differential
$\nabla^{\gamma_{C}}$. Formally, this implies that the connection
$\gamma_s$ is reducible to $\gamma_C$ (but in any case not
projectable), so the following diagram is commuting:
$$\commdiag{P\times S\otimes U(1)&\mapright^{\gamma_s}&J^1Y_h\cr
\mapup\lft{i_h}&&\mapup\rt{J^1i_h}\cr
Y_h&\mapright^{\gamma_C}&J^1P\times S\otimes U(1)\cr}$$ where the
inclusion map $i_h$ is the one of relation (\ref{8inc}) and
$J^1i_h$ the jet prolongation of this inclusion map.

\subsection{Geometric Structures for the Bosonic Sector}

The same arguments we employed for the fermionic sector, also hold
for the bosonic sector. In this case the geometric structure
implied by the $N=2$ SUSY QM underlying the bosonic sector, is
represented by the following diagram:
$$
\harrowlength=130pt \varrowlength=60pt \sarrowlength=120pt
\commdiag{X&\mapright^{\gamma_B}&TX^*\times \mathcal{C}\otimes
U(1){\,}{\,}{\,}{\,}{\,}\cr
&\arrow(3,-2)\lft{\gamma_E}&\mapup\rt{\gamma_{BE}}\cr &&
{\,}{\,}{\,}TX^*\otimes \wedge \mathcal{H_E}^*\otimes
\mathcal{H_E}\cr}$$ As in the fermionic sector, $X$ is the
$(2+1)$-dimensional manifold, while the $U(1)$-twisted fibre
bundle $TX^*\times \mathcal{C}\otimes U(1)$ has sections that are
actually the bosonic field variations $\delta \phi$, $\delta A_i$,
$i=1,2$. Hence, we have a composite fibre bundle
\begin{equation}\label{compbund1}
TX^*\times \mathcal{C}\otimes U(1)\xrightarrow{\pi_{BE}}
TX^*\otimes \wedge \mathcal{H_E}^*\otimes \mathcal{H_E}
\xrightarrow{\pi_{E}} X
\end{equation}
The connections of this composite fibre bundle are defined
similarly to those of the fermionic sector, and are of the
following form:
\begin{align}\label{mapconn1}
& \gamma_B: TX^*\times \mathcal{C}\otimes U(1)\rightarrow
J^1(TX^*\times \mathcal{C}\otimes U(1)),
\\ \notag&(\mathrm{Bundle}{\,}\mathrm{map},{\,}{\,}\pi_B:TX^*\times \mathcal{C}\otimes U(1)\rightarrow X)
\\ \notag
& \gamma_E: TX^*\otimes \wedge \mathcal{H_E}^*\otimes
\mathcal{H_E} \rightarrow J^1(TX^*\otimes \wedge
\mathcal{H_E}^*\otimes \mathcal{H_E})
\\ \notag & (\mathrm{Bundle}{\,}\mathrm{map},{\,}{\,}\pi_E:TX^*\otimes \wedge \mathcal{H_E}^*\otimes \mathcal{H_E}\rightarrow X) \\ \notag
& \gamma_{BE}: (TX^*\times \mathcal{C}\otimes U(1))_G\rightarrow
J^1((TX^*\times \mathcal{C}\otimes U(1))_G)
\\ \notag &(\mathrm{Bundle}{\,}\mathrm{map},{\,}{\,}\pi_{BE}: TX^*\times \mathcal{C}\otimes U(1)\rightarrow TX^*\otimes \wedge \mathcal{H_E}^*\otimes \mathcal{H_E})
\end{align}
The covariant differential corresponding to the connection
$\gamma_B$, which we denote $\nabla^{\gamma_{B}}$, when it acts to
integral sections of the fibre bundle $\pi_B:TX^*\times
\mathcal{C}\otimes U(1)\rightarrow X$, yields the differential
equations (\ref{selfdualfluctuat1fd}). In addition, the covariant
differential corresponding to the composite connection
$\gamma_F=\gamma_{BE}\circ \gamma_E$, which we denote
$\nabla^{\gamma_{F}}$, when it acts to integral composite sections
of the composite fibre bundle, yields the same set of equations.
This can only be true if the connection $\gamma_B$ is projectable
over $\gamma_F$. Note that the covariant differential
$\nabla^{\gamma_{B}}$ is a map of the form:
\begin{equation}\label{totcovdifcomp1}
\nabla^{\gamma_{B}}=\mathcal{D}_{\gamma_{B}}\circ J^1 s_B:
X\rightarrow TX^*\otimes V(TX^*\times \mathcal{C}\otimes U(1))
\end{equation}
with $s_B$ the corresponding integral section, while the covariant
differential $\nabla^{\gamma_{F}}$ is of the form:
\begin{equation}\label{totcovdifcomp12}
\nabla^{\gamma_{F}}=\mathcal{D}_{\gamma_{F}}\circ J^1 s_F:
X\rightarrow TX^*\otimes VY'
\end{equation}
In the above relation, $Y'$ is some subbundle of the fibre bundle
$TX^*\times \mathcal{C}\otimes U(1)$. The fact that $\gamma_{B}$
is projectable over (but not reducible) $\gamma_F$, means that the following diagram
is commuting:
$$\commdiag{TX^*\times \mathcal{C}\otimes U(1)&\mapright^{\gamma_B}&J^1(TX^*\times \mathcal{C}\otimes U(1))\cr
\mapdown\lft{\pi_{bs}}&&\mapdown\rt{J^1\pi_{bs}}\cr
Y'&\mapright^{\gamma_F}&J^1Y'\cr}$$ with $\pi_{bs}$, the
projection $\pi_{bs}:TX^*\times \mathcal{C}\otimes U(1)\rightarrow
Y'$.

\subsection{Bundle Isomorphisms Between Fermionic and Bosonic Sectors and K-theoretic Arguments}

Recall relations (\ref{susyqmrn567m}) and (\ref{susyqmrrtyurn567m}). The
operators $\mathcal{D}_{LG}$ and $\mathcal{D}_{LG}'$ apart from
the fact that they act in a subset of fermionic sections and
bosonic sections respectively, they are identical. This is very
important from a geometric aspect, in view of the results we
presented in the previous two subsections. Clearly, such an
equality is by far not accidental. Recall that these operators are
actually the covariant differentials of the corresponding
composite fibre bundles we saw earlier. Hence the covariant
differentials $\nabla^{\gamma_F}$ and $\nabla^{\gamma_B}$ are
actually equal. This fact clearly signals some sort of equivalence
between the fibre bundle $Y_h$ and $Y'$, which can be quantified
by the existence of an isomorphism $\Phi$, so that the following
diagram is commuting:
$$
\harrowlength=60pt \varrowlength=40pt \sarrowlength=60pt
\commdiag{Y'&\mapright^{\Phi}&Y_h\cr
&\arrow(3,-2)\lft{\pi_{b}}&\mapdown\rt{\pi_{s}}\cr && X\cr}$$ with
$\pi_b$ and $\pi_s$ the projections from the total bundle space to
the base space $X$ of the fibre bundles $Y'\rightarrow X$ and
$Y_h\rightarrow X$ respectively. We can establish that conclusion
by thinking as follows: Since $\nabla^{\gamma_F}\equiv
\nabla^{\gamma_B}$, this implies some isomorphism between the
fibre bundles $TX^*\otimes VY'$ and $TX^*\otimes VY_h$, which
isomorphism induces an isomorphism between the vertical subbundles
$VY'$ and $VY_h$. In turn, this isomorphism induces the
isomorphism between the spaces $Y'$ and $Y_h$ (recall that $Y'$
and $Y_h$ are affine vector bundles).

\noindent One could argue that this bundle equivalence between
$Y'$ and $Y_h$, could be extended to some sort of
K-theoretic equivalence. Indeed, following the same line of argument as
above, the bundles $Y_h\oplus I^n$ and $Y'\oplus I^n$, with $I^n$
some trivial bundle over $X$, are stably equivalent, that is:
\begin{equation}\label{sequiv}
Y_h\oplus I^n\approx Y'\oplus I^n
\end{equation}
Hence they belong to the same equivalence class in $K(X)$, that is
$[Y']=[Y_h]$.

\section*{Concluding Remarks}

In this paper we studied an abelian $N=2$ supersymmetric
extension of the Landau-Ginzburg
model. Both the bosonic and the fermionic sectors
have a common underlying unbroken $N=2$ SUSY QM algebra. This structure
provides both sectors with an additional geometric structure over
the $(2+1)$-dimensional space $X$. Particularly, this geometric
structure is based on a composite bundle over the space $X$, with
a graded manifold over $X$, being a basic ingredient of the
composite bundle. We saw that this geometric structure underlies
both the fermionic and the bosonic sector, with the difference
that in the bosonic sector, the covariant differential of the
corresponding total bundle $\nabla^{\gamma_{B}}$ is projectable to $\nabla^{\gamma_{F}}$ corresponding
to the total bundle space that is related to the graded manifold.  In the fermionic case, the
covariant differential $\nabla^{\gamma_{s}}$ is reducible to $\nabla^{\gamma_{C}}$
corresponding to some subbundle $Y_h$ of the initial total twisted spin
bundle.

\noindent Due to the fact that the zero modes of the fermions and
bosons are related, the geometric structures corresponding to
bosons and fermions are equivalent. In particular, we found a
direct correlation between the fermionic and bosonic bundles, in
terms of an isomorphism. This result is a consequence of the
global $N=2$ supersymmetry that the system possesses and owing to the fact that the space $X$ is $(2+1)$-dimensional. It
is intriguing that, apart from the $N=2$ supermanifold we can construct
over $X$, we also found that $X$ is a $Z_2$-graded manifold, with
the grading being provided by the involution $W$, which is the
Witten parity of the SUSY QM system. It would be interesting to
try to find if there is any direct correlation between the
supermanifold corresponding to global $N=2$ supersymmetry and that of the graded manifold (which is not a
supermanifold). This task is highly motivated by the fact that every
graded manifold defines a DeWitt supermanifold. However, we must
be cautious because the SUSY QM in the bosonic sector is related
to the bosonic fluctuations. This would require analysis of some
corresponding jet module, something that is interesting but also
out of the scope of this paper.

\noindent Finally, in order to support our arguments for the constructed geometric structures, we used only integral sections of 
the corresponding fibre bundles. The integral sections are very useful for the quantum theory of Chern-Simons vortices. This is due to the fact that the bosonic ones
correspond to collective coordinates describing the vortices positions and slow velocity kinematics. Moreover, the fermionic ones represent the degeneracy of the solitonic states.
Hence, we believe that this extra geometric structure of the SUSY Chern-Simons model, is relevant for the complete quantum theory of Chern-Simons vortices.


\begin{thebibliography}{99}

\bibitem{lee1} R. Jackiw and S. Templeton, Phys. Rev. D 23, 2291
(1981)

\bibitem{lee1a} J. Schonfeld, Nucl. Phys. B185, 157 (1981)

\bibitem{lee1b} S.
Deser, R. Jackiw, and S. Templeton, Phys. Rev. Lett. 48, 975
(1982)

\bibitem{dunnebook} Gerald Dunne, ``Self-Dual Chern-Simons Theories'', Springer, Berlin 1995; S.D. Odintsov, A.V. Timoshkin, Russ.Phys.J. 35 (1992) 529-533; S.D. Odintsov, Z.Phys. C54 (1992) 527-530

\bibitem{lee} Bum-Hoon Lee, C. Lee, H. Min, Phys. Rev. D 45, 4588 (1992)

\bibitem{workinprogress} V.K. Oikonomou, Nucl. Phys. B, 870, 477 (2013)



\bibitem{witten} E. Witten, Nucl. Phys. B249, 557 (1985)

\bibitem{susyqm}  G. Junker, ``Supersymmetric Methods in Quantum and Statistical Physics", Springer, 1996

\bibitem{susyqm1}M. Combescure, F. Gieres, M. Kibler, J. Phys.
A: Math. Gen. 37, 10385 (2004)

\bibitem{susyqm2} Asim Gangopadhyaya, Jeffry V Mallow, Constantin Rasinariu, ''Supersymmetric Quantum Mechanics, An Introduction'', World Scientific

\bibitem{susyqm3} A. Kirchberg, J.D. Laenge, A. Wipf, Annal.Phys., 315, 467 (2005)

\bibitem{susyqm4} F. Cooper, A. Khare, U. Sukhatme, Physics Reports, 251, 267 (1995)

\bibitem{susyqm5}  Jonathan Bougie 1, Asim Gangopadhyaya, Jeffry Mallow, Constantin Rasinariu, Symmetry 4(3), 452 (2012)



\bibitem{susyqmarxiv}
 Zhanna Kuznetsova , Francesco Toppan, Int. J. Mod. Phys. A23 (2008) 3947


\bibitem{susyqmarxiv1} A. S. Galperin, E. A. Ivanov, V. I. Ogievetsky, E. S. Sokatchev, `` Harmonic Superspace'', Cambridge Monographs on Mathematical Physics,  2007;  C.M. Hull, hep-th/9910028; F. Delduc, S. Kalitzin, E. Sokatchev, Class. Quantum Grav. 7 (1990) 1567; E. Ivanov, O. Lechtenfeld, A. Sutulin,  Nucl. Phys.
B 790 (2008) 493; E. Ivanov, Int.J.Geom.Meth.Mod.Phys. 09, 1261006  (2012)
 
\bibitem{susyqmarxiv2}

V.P. Berezovoi, A.I. Pashnev, Theor.Math.Phys. 74 (1988) 264, Teor.Mat.Fiz. 74 (1988) 392


\bibitem{susyqmarxiv} Alexander A. Andrianov, N.V. Borisov, Mikhail V. Ioffe, Michael I. Eides, Theor.Math.Phys. 61 (1984) 965, Teor.Mat.Fiz. 61 (1984) 17; J. Gegelia, L.A. Slepchenko, Theor.Math.Phys. 86 (1991) 50, Teor.Mat.Fiz. 86N1 (1991) 74; A.I. Pashnev, Theor.Math.Phys. 69 (1986) 1172; Teor.Mat.Fiz. 69 (1986) 311; N. Berezovoi, A.I. Pashnev, Theor.Math.Phys. 70 (1987) 102, Teor.Mat.Fiz. 70 (1987) 146

\bibitem{susyqmarxiv3}

 B. Bagchi, S. Mallik, C. Quesne, Int. J. Mod. Phys. A16 (2001) 2859; V. I. Tkach, Pashnev A. I., Rosales J. J., Mod.Phys.Lett. A15 (2000) 1557

\bibitem{susybreaking} V.P. Akulov, A.I. Pashnev, Theor.Math.Phys. 65 (1985) 1027; E. Witten, Int.J.Mod.Phys.A10:1247 (1995)


\bibitem{susyqmarxiv4}

 D. Spector, Int.J.Mod.Phys. A20 (2005) 6288;  D. Ruan, C.C. Tu, H.Z. Sun, Commun.Theor.Phys. 32 (1999) 477;  C. Quesne, Mod.Phys.Lett. A18 (2003) 515 E.A. Ivanov, S.O. Krivonos, V.M. Leviant, Int.J.Mod.Phys. A7 (1992) 287-316





\bibitem{thaller} ``The Dirac Equation", Bernd Thaller, Springer 1992

\bibitem{pluskai}  M. S. Plyushchay, Annals
Phys.245, 339 (1996); Francisco Correa
Mikhail S. Plyushchay, Annals Phys. 322, 2493 (2007)




\bibitem{plu1}  Adrian Arancibia, Juan Mateos Guilarte, Mikhail S. Plyushchay, Phys.Rev. D87 (2013) 4, 045009; Alberto Alonso-Izquierdo, Juan Mateos Guilarte, Mikhail S. Plyushchay, Annals Phys. 331 (2013) 269; Mikhail S. Plyushchay, Adrian Arancibia, Luis-Miguel Nieto.
[arXiv:1012.4529, Phys.Rev. D83 (2011) 065025; Adrian Arancibia, Mikhail S. Plyushchay, Phys.Rev. D85 (2012) 045018; 


\bibitem{plu2}Francisco Correa, Vit Jakubsky, Luis-Miguel Nieto, Mikhail S. Plyushchay, Phys.Rev.Lett. 101 (2008) 030403; Francisco Correa, Vit Jakubsky, Mikhail S. Plyushchay, J.Phys. A41 (2008) 485303; Francisco Correa, Vit Jakubsky, Mikhail S. Plyushchay, 
Annals Phys. 324 (2009) 1078; 


\bibitem{plu3}Francisco Correa, Luis-Miguel Nieto, Mikhail S. Plyushchay, Phys.Lett. B659 (2008) 746; Sergey M. Klishevich, Mikhail S. Plyushchay, Nucl.Phys. B606 (2001) 583-612

\bibitem{plu4}  Mikhail Plyushchay, Int.J.Mod.Phys. A15 (2000) 3679-3698


\bibitem{ivanovbook} A. S. Galperin, E. A. Ivanov, V. I. Ogievetsky, E. S. Sokatchev, `` Harmonic Superspace'', Cambridge Monographs on Mathematical Physics,  2007


\bibitem{ivanov1} F. Delduc, E.A. Ivanov, Nucl. Phys. B
855 (2012) 815, arXiv:1107.1429; E. Ivanov, Int.J.Geom.Meth.Mod.Phys. 09, 1261006  (2012); E.A. Ivanov, S.O. Krivonos, A.I. Pashnev, Class.Quant.Grav. 8 (1991) 19-40 

\bibitem{ivanov2} C.M. Hull, The Geometry of Supersymmetric Quantum Mechanics, hep-th/9910028

\bibitem{ivanov3} E.A. Ivanov, A.V. Smilga, Dirac Operator on Complex Manifolds and Supersymmetric
Quantum Mechanics, arXiv:1012.2069

\bibitem{ivanov4} A. Galperin, E. Ivanov, V. Ogievetsky, E. Sokatchev, Hyper-Kaehler metrics and harmonic
superspace, Commun. Math. Phys. 103 (1986) 515

\bibitem{ivanov5} F. Delduc, S. Kalitzin, E. Sokatchev, Geometry of sigma models with heterotic supersymmetry,
Class. Quantum Grav. 7 (1990) 1567

\bibitem{ivanov6} A.S. Galperin, E.A. Ivanov, V.I. Ogievetsky and E.S. Sokatchev, Harmonic Superspace,
Cambridge University Press 2001

\bibitem{ivanov7} E. Ivanov, O. Lechtenfeld, N=4 supersymmetric mechanics in harmonic superspace,
JHEP 0309 (2003) 073, hep-th/0307111.

\bibitem{ivanov8} E. Ivanov, O. Lechtenfeld, A. Sutulin, Hierarchy of N=8 Mechanics Models, Nucl. Phys.
B 790 (2008) 493, arXiv:0705.3064

\bibitem{ivanov9} F. Delduc, E. Ivanov, arXiv: 1201.3794


\bibitem{ivanov10} Curtright T., Mezincescu L., Ivanov E., Townsend P.K., Planar super-Landau models revisited,
JHEP, 020, 25 (2007), hep-th/0612300


\bibitem{ivanov11} Ivanov E., Mezincescu L., Townsend P.K., JHEP 143, 23, (2006), hep-th/0510019



\bibitem{graded1}G. Sardanashvily, Theor.Math.Phys. 114, 368 (1998), Teor.Mat.Fiz. 114 (1998) 470-480;  G. Sardanashvily, O. Zakharov, Gauge Gravitation Theory, World Scientific  (1992); G.  Sardanashvily, Generalized Hamiltonian Formalism for Field Theory, World Scientific, (1995); G. Giachetta, L. Mangiarotti, G. Sardanashvily, New Lagrangian and Hamiltonian Methods in Field Theory, World Scientific (1997); L. Mangiarotti, G.  Sardanashvily, Gauge Mechanics, World Scientific, (1998); G. Giachetta, L. Mangiarotti, G. Sardanashvily,  Geometric and Algebraic Topological Methods in Quantum Mechanics, World Scientific  (2005); G. Giachetta, L. Mangiarotti, G. Sardanashvily,  Geometric Methods in Classical and Quantum Mechanics, World Scientific  (2010)

\bibitem{Jost} ``Riemannian Geometry and Geometric Analysis",
J. Jost, Universitext, Springer, 2005

\bibitem{Nakahara} ``Geometry, Topology and Physics", M. Nakahara, Graduate Student Series in Physics, IOP Publishing, 1990

\bibitem{eguchi} T. Eguchi, P. B. Gilkey, and A. J. Hanson, Phys. Rept. 66, 213
(1980)

\bibitem{graded} T. Stavracou, Rev. Math. Phys. 10, 47 (1998)













\end{thebibliography}
\end{document}